\documentclass[11pt]{article}
\usepackage{ae}
\usepackage[T1]{fontenc}
\usepackage[nottoc]{tocbibind}
\usepackage{amsmath}
\usepackage{amsthm}
\usepackage{amssymb}
\usepackage{braket} %usato per \set e \Set
\usepackage{mathtools} %usato per \coloneqq e \DeclarePairedDelimiter
\usepackage{dsfont} %usato per \mathds
\usepackage{accents} %usato per \accentset
\usepackage[all]{xy}
\usepackage{booktabs}
\usepackage{slashbox}

\SelectTips{cm}{} %imposta lo stile delle frecce per il pacchetto XY-pic

\makeatletter
\let\origthm@notefont\thm@notefont
\renewcommand{\thm@notefont}[1]{\origthm@notefont{#1\the\thm@headfont}}
\makeatother

\newtheorem{thm}{Theorem}[section]
\newtheorem{theorem}[thm]{Theorem}

\newtheorem{lemma}[thm]{Lemma}
\newtheorem{proposition}[thm]{Proposition}

\theoremstyle{definition}
\newtheorem{definition}[thm]{Definition}
\newtheorem{example}[thm]{Example}

\newtheorem{remark}[thm]{Remark}

\newcommand{\cat}[1]{{\boldsymbol{\mathsf{#1}}}} %categoria
\newcommand{\sets}{\cat{Set}}
\newcommand{\grp}{\cat{Grp}}

\newcommand{\smflds}{\cat{SMan}}
\newcommand{\cfunct}[2]{\left[{#1}\op,{#2}\right]} %categoria dei funtori controvarianti tra due categorie
\newcommand{\yo}{\mathcal Y} %Yoneda embedding

\newcommand{\nset}[1]{{\mathds{#1}}} %insieme numerico

\newcommand{\Z}{\nset{Z}}
\newcommand{\R}{\nset{R}}

\renewcommand{\to}{\mathchoice{\longrightarrow}{\rightarrow}{\rightarrow}{\rightarrow}} %\to (corto nel testo e lungo nelle equazioni)
\let\mto\mapsto \renewcommand{\mapsto}{\mathchoice{\longmapsto}{\mto}{\mto}{\mto}} %\mapsto (corto nel testo e lungo nelle equazioni)

\newcommand{\id}{\mathds{1}} %identita`
\newcommand{\pr}{\mathrm{pr}} %proiezione
\DeclareMathOperator{\im}{Im} %immagine
\newcommand{\isom}{\cong} %isomorfo

\newcommand{\Hom}{\mathrm{Hom}}
\newcommand{\HOM}{\underline{\Hom}}

\newcommand{\restr}[2]{{#1}_{|{#2}}} %restrizione

\newcommand{\gk}{\mathfrak{g}}
\newcommand{\pk}{\mathfrak{p}}
\newcommand{\hk}{\mathfrak{h}}

\newcommand{\gl}{\mathfrak{gl}}
\newcommand{\Uk}{\mathfrak{U}} %universal enveloping algebra

\newcommand{\sheaf}[1][O]{\mathcal{#1}} %fascio
\newcommand{\cinfty}{\sheaf[C]^\infty} %smooth things
\newcommand{\hotimes}{\mathbin{\widehat{\otimes}}} %algebraic tensor product
\newcommand{\topored}[1]{|{#1}|} %ridotto/a; es.: \red{x_A}
\newcommand{\red}[1]{\widetilde #1} %ridotto/a; es.: \red{x_A}
\newcommand{\lgr}[1]{#1_0} %gruppo classico super copia: \lgr{x_A}
\newcommand{\VEC}{\mathrm{Vec}} %campi vettoriali
\newcommand{\ev}{\mathrm{ev}} %valutazione

\newcommand{\p}[1]{|#1|} %parita` (potrebbe essere impostato come o $p(x)$ o $|x|$)
\newcommand{\op}{^{\mathrm{op}}} %opposto
\newcommand{\blank}{\mspace{2mu}{\cdot}\mspace{2mu}} %puntino segnaposto
\newcommand{\Lie}{\mathrm{Lie}}
\newcommand{\Gl}{\mathrm{Gl}}
\newcommand{\ua}{\underline{a}}

\newcommand{\nbd}{\nobreakdash-\hspace{0pt}}

\renewcommand{\phi}{\varphi}
\renewcommand{\theta}{\vartheta}
\renewcommand{\epsilon}{\varepsilon}

\begin{document}
\centerline{\Large\bf Super $G$-spaces}
\bigskip

\centerline{L. Balduzzi$^\natural$, C. Carmeli$^\natural$, G.
Cassinelli$^\natural$}

\bigskip

\centerline{\it $^\natural$Dipartimento di Fisica, Universit\`a di
Genova and INFN, sezione di Genova} \centerline{\it Via Dodecaneso,
33 16146 Genova, Italy} \centerline{\footnotesize e-mail:
luigi.balduzzi@ge.infn.it, claudio.carmeli@ge.infn.it,
cassinelli@ge.infn.it}

\begin{abstract}
We review the basic theory of super $G$-spaces. We prove a theorem
relating the action of a super Harish-Chandra pair $(\lgr{G}, \gk)$
on a supermanifold to the action of the corresponding super Lie
group $G$. The theorem was stated in \cite{DM} without proof. The
proof given here does not use Frobenius theorem but relies on Koszul
realization of the structure sheaf of a super Lie group (see
\cite{Koszul}). We prove the representability of the stability
subgroup functor.
\end{abstract}

\tableofcontents

\section{Introduction}

In his seminal paper \cite{Kostant}, B. Kostant gave a complete and
rigorous foundation of supergeometry, including super Lie groups. He
introduced, for the first time, the notion of super Harish-Chandra
pair (called Lie--Hopf algebra, in that paper) and proved the
equivalence between those and super Lie groups (see also \cite{DM},
where the name super Harish-Chandra pair was introduced).

In this paper we review the basic aspects of the theory of smooth
actions of super Lie groups on supermanifolds. The language we adopt
is different than that used by Kostant in \cite{Kostant}. In
particular we use the explicit realization of the sheaf of a  super
Lie group in terms of the corresponding super Harish-Chandra pair,
as given by Koszul in \cite{Koszul}. This has the advantage that
many constructions become more transparent and easy to prove.

In the first sections we briefly recall the basic definitions  and
results on super Lie groups and super Harish-Chandra pairs. In
particular we state the precise link existing between them giving an
explicit construction of the equivalence of the two categories. This
is the main ingredient of all subsequent results. In section
\ref{sec:G-superman}, we recall the concept of action of a super Lie
group $G$ on a supermanifold $M$, and in prop.\
\ref{prop:action_SLG_SHCP} we establish the precise link between
super Lie group actions and super Harish-Chandra pair actions (the
proposition was stated without proof in \cite{DM}). In section
\ref{sec:trans_action}, the notion of transitive action is analyzed
and characterized  both from the point of view of super
Harish-Chandra pairs and from the point of view of the functor of
points. In section \ref{sec:stabilizer} we consider the stabilizer
of a supergroup action and a representability theorem for the
stability group functor is given. Finally in the last section we
review the construction of super homogeneous spaces.

\section{Supermanifolds and super Lie groups}

A \emph{supermanifold} $M$ is a locally compact, second countable,
Hausdorff topological space $\topored{M}$ endowed with a sheaf
$\sheaf_M$ of superalgebras, locally isomorphic to $\cinfty(\R^p)
\otimes \Lambda(\theta_1,\ldots,\theta_q)$. A morphism $\psi \colon
M \to N$ between supermanifolds  is a pair of morphisms
$(\topored{\psi}, \psi^\ast)$ where $\topored{\psi} \colon
\topored{M}\to\topored{N}$ is a continuous map and $\psi^\ast \colon
\sheaf_N \to \sheaf_M$ is a sheaf morphism above $\topored{\psi}$.

\begin{remark}
We will consider only smooth supermanifolds. It can be proved that
in this category a morphism of supermanifolds is determined once we
know the corresponding morphism of the global sections (see, for
example, \cite{Kostant} and \cite{BBHR}). In other words, a morphism
$\psi \colon M \to N$ can be identified with a superalgebra map
$\psi^* \colon \sheaf_N(\topored{N}) \to \sheaf_M(\topored{M})$. We
will tacitly use  this fact several times. Moreover,  in the
following, we will denote with $\sheaf(M)$ the superalgebra of
global sections $\sheaf_M(\topored{M})$.
\end{remark}

Suppose now $U$ is an open subset of $\topored{M}$ and let
$\sheaf[J]_M(U)$ be the ideal generated by the nilpotent elements of
$\sheaf_M(U)$. It is possible to prove that $\sheaf_M/\sheaf[J]_M$
defines a sheaf of purely even algebras over $\topored{M}$ locally
isomorphic to $\cinfty(\R^p)$. Therefore $\red{M} \coloneqq
(\topored{M},\sheaf_M/\sheaf[J]_M)$ defines a classical manifold,
called the \emph{reduced manifold} associated to $M$. Analogously it
is possible to prove that each supermanifold morphism $\psi \colon
M\to N$ determines a corresponding \emph{reduced map} $\red{\psi}
\colon \red{M} \to \red{N}$. The map $f \mapsto \red{f} \coloneqq f+
\sheaf[J]_M(U)$, with $f \in \sheaf_M(U)$, defines the embedding
$\red{M} \to M$. In the following we will denote with $\ev_p(f)
\coloneqq \red{f}(p)$ the evaluation at $p \in U$.

An important and very used tool in working with supermanifolds is
the functor of points. Given a supermanifold $M$ one can construct
the functor $M(\blank) \colon \smflds\op \to \sets$ from the
opposite of the category of supermanifolds to the category of sets
defined by $S \mapsto M(S) \coloneqq \Hom(S,M)$ and called the
\emph{functor of points} of $M$. In particular, for example,
$M(\R^{0|0}) \isom \topored{M}$ as sets. Each supermanifold morphism
$\psi \colon M \to N$ defines the natural transformation
$\psi(\blank) \colon M(\blank) \to N(\blank)$ given by $[\psi(S)](x)
\coloneqq \psi \circ x$. Due to Yoneda's lemma, each natural
transformation between $M(\blank)$ and $N(\blank)$ arises from a
unique morphism of supermanifolds in the way just described. The
category of supermanifolds can thus be embedded into a full
subcategory of the category $\cfunct{\smflds}{\sets}$ of functors
from the opposite of the category of supermanifolds to the category
of sets. Let
\[
    \begin{aligned}
        \yo \colon \smflds &\to \cfunct{\smflds}{\sets}\\
        M &\mapsto M(\blank)
    \end{aligned}
\]
denote such embedding. It is a fact that the image of $\smflds$
under $\yo$ is strictly smaller than  $\cfunct{\smflds}{\sets}$. The
elements of $\cfunct{\smflds}{\sets}$ isomorphic to elements in the
image of $\yo$ are called \emph{representable}. Supermanifolds can
thus be thought as the representable functors in
$\cfunct{\smflds}{\sets}$. For all the details we refer to
\cite{Kostant, Leites, Manin, DM, Varadarajan}.

\bigskip

\emph{Super Lie groups (SLG)} are, by definition, group objects in
the category of supermanifolds. This means that morphisms $\mu$, $i$
and $e$ are defined satisfying the usual commutative diagrams for
multiplication, inverse and unit respectively. From this, it follows
easily that the reduced morphisms $\red{\mu}, \red{i}$, and
$\red{e}$ endow $\red{G}$  with a Lie group structure. $\red{G}$ is
called the \emph{reduced (Lie) group} associated with $G$. $\red{G}$
acts in a natural way on $G$. In particular, in the following, we
will  denote by
\begin{align*}
    r_g &\coloneqq \mu \circ \langle \id_G,\hat{g} \rangle &
    \ell_g &\coloneqq \mu \circ \langle \hat{g}, \id_G \rangle
\end{align*}
the right and left translations by the element
$g\in\red{G}$, respectively\footnote{%
    \label{footnote:notations}%
    Some explanations of the notations used: given two morphisms
    $\alpha \colon X \to Y$ and $\beta \colon X \to Z$,
    \[
        \langle \alpha,\beta \rangle \colon X \to Y \times Z
    \]
    is the morphism that composed with the projection on the first
    (resp.\ second) component gives $\alpha$ (resp.\ $\beta$); if $x \in \red{X}$, the map
    \[
        \hat{x} \colon T \to X
    \]
    is the constant map obtained composing the unique map $T \to \R^{0|0}$
    with the embedding $\R^{0|0} \to X$ whose image is $x$.
}.

Many classical constructions carry over to the super setting. For
example it is possible to define \emph{left\nbd invariant vector
fields} and to prove that they form a super Lie algebra $\gk$,
isomorphic to the super tangent space at the identity of $G$ (see,
for example, \cite{Kostant} or \cite{Varadarajan}).

In the spirit of the functor of points, one can think of a SLG as a
representable  functor from $\smflds\op$ to the category $\grp$ of
set theoretical groups. The SLG structure imposes severe
restrictions on the structure of the supermanifold carrying it. In
the next section, we want to briefly discuss this point.

\section{Super Harish-Chandra pairs}

\begin{definition} \label{def:SHCP}
Suppose $(\lgr{G},\gk,\sigma)$ are respectively a Lie group, a super
Lie algebra and a representation of $\lgr{G}$ on $\gk$ such that
\begin{enumerate}
    \item $\gk_0 \isom \Lie(\lgr{G})$,
    \item $\restr{\sigma}{\gk_0}$ is equivalent to the adjoint representation of $\lgr{G}$ on
    $\gk_0$.
\end{enumerate}
$(\lgr{G},\gk,\sigma)$ is called a \emph{a super Harish-Chandra pair
(SHCP)}.
\end{definition}

\begin{example}
Let $G$ be a SLG, it is clear that we can associate to it the SHCP
given by:
\begin{enumerate}
    \item the reduced Lie group $\red{G}$;
    \item the super Lie algebra $\gk=\gk_0\oplus \gk_1$
    of $G$; notice that $\gk_0 \isom \Lie \big( \red{G} \big)$;
    \item the adjoint representation of $\red{G}$ on $\gk$ given by\footnote{%
        If $M$ is a supermanifold and $U$ is an open subset of $\topored{M}$ we endow $\sheaf_M(U)$ with
        the usual topology considered in \cite{Kostant}.  As in the
        classical case (see for example \cite{Grothendieck}), it can be
        proved that if $M$ and $N$ are two supermanifolds and $U \times V
        \subseteq M \times N$ is an open subset, $\sheaf_{M\times N}(U\times
        V)$  can be identified with the completed projective tensor product
        $\sheaf_M(U) \hotimes \sheaf_N(V)$. This fact will be used each time we
        will write a morphism between supermanifolds in the tensor product
        form. Moreover since all the maps we will consider are continuous in
        the given topology, we will check formulas only on decomposable
        elements. The reader can easily work out the details each time.
    }
    \[
        \mathrm{Ad}(g)X \coloneqq (\ev_g \otimes X \otimes \ev_{g^{-1}})(\id \otimes \mu^*) \mu^*
    \]
    with $g \in \red{G}$ and $X \in \gk$ ($X$ is thought as a vector in $T_e(G)$).
\end{enumerate}
\end{example}

\begin{definition} \label{def:SHCP_morph}
If $(\lgr{G},\gk,\sigma)$ and $(\lgr{H},\hk,\tau)$ are SHCP, a
morphism between them is a pair of morphisms
\begin{align*}
    \psi_0 \colon \lgr{G} &\to \lgr{H} \\
    \rho_\psi \colon \gk &\to \hk
\end{align*}
satisfying the compatibility conditions
\begin{enumerate}
    \item $\restr{\rho_\psi}{\gk_0} \isom d\psi_0$;
    \item $\rho_\psi \circ \sigma(g) = \tau\big( \psi_0(g) \big) \circ
    \rho_\psi$ for all $g \in \lgr{G}$.
\end{enumerate}
\end{definition}

\begin{example}
If $\psi \colon G \to H$ is a SLG morphism, then it defines  the
morphism between the associated SHCP given by $\psi_0 = \red{\psi}$
and $\rho_\psi = d\psi$.
\end{example}

Definitions \ref{def:SHCP} and \ref{def:SHCP_morph} allow to define
the category $\cat{SHCP}$ of super Harish-Chandra pairs. Moreover
the above examples show that the correspondence
\begin{equation} \label{eq:functor}
    \begin{aligned}
        \cat{SGrp} &\to \cat{SHCP} \\
        G &\mapsto \big( \red{G}, \Lie(G), \mathrm{Ad} \big)
    \end{aligned}
\end{equation}
is functorial.

The following is a crucial result in the development of the theory.
\begin{theorem}[B. Kostant]
The functor \eqref{eq:functor} defines an equivalence of categories.
\end{theorem}

It is fundamental to notice that it is possible to give a very
explicit form to the inverse functor (see Koszul's paper
\cite{Koszul}). We now want to briefly describe it.

Let us preliminarily remember that the super enveloping algebra
$\Uk(\gk)$ can be endowed with a super Hopf algebra structure (see
Kostant's paper \cite{Kostant}). In fact it is a unital superalgebra
with respect to the natural identity $1_{\Uk(\gk)}$ and
multiplication $m_{\Uk(\gk)}$. Moreover the map $\gk \to \gk \otimes
\gk$ defined by $X\mapsto X\otimes 1 + 1\otimes X$ can be extended
to a comultiplication map
\[
    \Delta_{\Uk(\gk)} \colon \Uk(\gk) \to \Uk(\gk)\otimes \Uk(\gk)
\]
in such a way that together with the counit
\[
    \epsilon \colon \Uk(\gk) \to \R
\]
$\Uk(\gk)$  becomes a super bialgebra. The antipode is finally
defined as the super antiautomorphism
\[
    \begin{aligned}
        S \colon \Uk(\gk) &\to \Uk(\gk) \\
        X &\mapsto \overline{X}
    \end{aligned}
\]
whose action on $\gk$ is given by $X \mapsto -X$. Clearly
$\overline{XY}=(-1)^{\p{X}\p{Y}}\overline{Y}\,\overline{X}$.

Suppose hence a SHCP $(\lgr{G},\gk,\sigma)$ is given and notice that
\begin{enumerate}
    \item $\Uk(\gk)$ is naturally a left $\Uk(\gk_0)$-module;
    \item $\cinfty(\lgr{G})$ is a left $\Uk(\gk_0)$ module. In fact each
    $X\in\Uk(\gk_0)$ acts from the left on smooth functions on $G_0$ as
    the left invariant differential operator $\red{D^L_X}$.
\end{enumerate}
Hence, for each open subset $U \subseteq \lgr{G}$, it is meaningful
to consider\footnote{%
    We recall that if $V = V_0 \oplus V_1$ and
    $W = W_0 \oplus W_1$ are super vector spaces then $\HOM(V,W)$
    denotes the super vector space of all linear
    morphisms between $V$ and $W$ with the gradation $\HOM(V,W)_0
    \coloneqq \Hom(V_0,W_0) \oplus \Hom(V_1,W_1)$, $\HOM(V,W)_1
    \coloneqq \Hom(V_0,W_1) \oplus \Hom(V_1,W_0)$.
}
\[
    \sheaf_G(U) \coloneqq \HOM_{\Uk(\gk_0)} \big( \Uk(\gk), \cinfty(U) \big)
\]
where the r.~h.~s.\ is the subset of $\HOM \big( \Uk(\gk),
\cinfty(U) \big)$ consisting of $\Uk(\gk_0)$-linear morphisms.

$\sheaf_G(U)$ has a natural structure of unital, commutative
superalgebra. The multiplication $\sheaf_G(U) \otimes \sheaf_G(U)
\to \sheaf_G(U)$ is defined by
\begin{equation} \label{eq:multiplication}
    \phi_1 \cdot \phi_2 \coloneqq m_{\cinfty(\lgr{G})} \circ (\phi_1 \otimes \phi_2) \circ \Delta_{\Uk(\gk)}
\end{equation}
and the unit is (with a mild abuse of notation) $\epsilon$.

The following proposition and lemma  are stated in Koszul's paper
\cite{Koszul}.

\begin{proposition}
$(\lgr{G}, \sheaf_G)$ is a supermanifold that is globally split
i.~e.
\begin{equation} \label{eq:global_split}
    \sheaf_G(\lgr{G})
    \isom \HOM \big(\Lambda(\gk_1),\cinfty(\lgr{G})\big)
    \isom \cinfty(\lgr{G}) \otimes {\Lambda(\gk_1)}^\ast
\end{equation}
$\sheaf_G$  carries a natural $\Z$-gradation.
\end{proposition}

The first isomorphism in \eqref{eq:global_split} is given by
\begin{equation} \label{eq:global_split_realiz}
    \phi \mapsto \phi \circ \gamma
\end{equation}
where $\gamma$ is the map defined in the following useful lemma,
that will be needed also in the following.

\begin{lemma} \label{lemma:antisym}
\begin{itemize}
    \item The antisymmetrizer
    \[
        \begin{aligned}
            \gamma \colon \Lambda(\gk_1) &\to \Uk(\gk) \\
            X_1 \wedge \cdots \wedge X_p &\mapsto
            \frac{1}{p!} \sum_{\tau\in{\mathfrak{S}_p}} (-1)^\tau X_{\tau(1)} \cdots X_{\tau(p)}
        \end{aligned}
    \]
    is a super coalgebra morphism.

    \item The map
    \[
        \begin{aligned}
            \widehat{\gamma} \colon \Uk(\gk_0) \otimes \Lambda(\gk_1) &\to \Uk(\gk) \\
            X \otimes Y &\mapsto X \cdot \gamma(Y)
        \end{aligned}
    \]
    is an isomorphism of super left $\Uk(\gk_0)$-modules.
\end{itemize}
\end{lemma}

Next proposition exhibits explicitly the structure of a SLG in terms
of the corresponding SHCP.

\begin{proposition}
$(\lgr{G},\sheaf_G)$ is a SLG with respect to the operations
\begin{subequations} \label{eqs:grp_operations}
\begin{align}
    \big[ \mu^\ast (\phi) (X,Y) \big] (g,h) &= \big[ \phi \big( (h^{-1}.X) Y \big) \big](gh) \\
    \label{eq:inv}
    \big[ i^\ast (\phi) (X) \big] (g^{-1}) &=  \big[ \phi(g^{-1}.\overline{X}) \big](g) \\
    e^\ast(\phi) &=  \big[ \phi(1) \big](e)
\end{align}
\end{subequations}
where $X,Y \in \Uk(\gk)$, $g,h \in \lgr{G}$, $e$ is the unit of
$\lgr{G}$ and $g.X \coloneqq \sigma(g) X$. Moreover the associated
SHCP is precisely $(\lgr{G},\gk,\sigma)$.
\end{proposition}

In this approach  the reconstruction of a SLG morphism from a SHCP
one is very natural. Suppose indeed that $(\psi_0, \rho_\psi)$ is a
morphism from $(\lgr{G},\gk)$ to $(\lgr{H},\hk)$, and suppose
$\phi\in\sheaf_H(U)$. It is natural to define $\psi^\ast(\phi)$
through the following diagram
\[
    \xymatrix@C=12ex{
        \Uk(\gk) \ar^{\rho_\psi}[r] \ar@{-->}[d]_{\psi^\ast(\phi)} & \Uk(\hk) \ar[d]^{\phi}\\
        \cinfty_{\lgr{G}} \big( \psi_0^{-1}(U) \big) & \cinfty_{\lgr{H}}(U) \ar[l]^{\psi_0^*}
    }
\]
It is not difficult to prove that this defines a SLG morphism with
associated SHCP morphism $(\psi_0, \rho_\psi)$.

Let us finally collect a glossary of some frequently used operations
in Koszul realization, completing those given in eq.\
\eqref{eq:multiplication} and \eqref{eqs:grp_operations} (notice
that, since $(-1)^{\p{X}(\p{\phi} + \p{Y})} \phi(YX) = (-1)^{\p{X}}
\phi(YX)$, it is possible to slightly  simplify the form of some
expressions).

\medskip
\begin{center}
\begin{tabular}{@{} cc @{}}
    \toprule
    operation & formula \\
    \midrule
    evaluation map & $\red{\phi}=\phi(1)$ \\[1ex]
    left translation & $\big[ \ell_h^\ast(\phi) \big](X) = \red{\ell_h^\ast} \big( \phi(X) \big)$ \\[1ex]
    right translation &  $\big[ r_h^\ast(\phi) \big](X) = \red{r_h^\ast} \big( \phi(h^{-1}.X) \big)$ \\[1ex]
    left invariant vector fields & $(D^L_X\phi)(Y) = (-1)^{\p{X}} \phi(YX)$ \\[1ex]
    right invariant vector fields & $\big[ (D^R_X\phi)(Y) \big](g)= (-1)^{\p{X}\p{\phi}} \phi \big( (g^{-1}.X) Y \big)(g)$ \\
    \bottomrule
\end{tabular}
\end{center}
\medskip

\begin{example} \label{example:GL(1|1)}
We consider the SLG $G = \Gl(1|1)$. Formally it can be thought as
the set of invertible matrices $
    \left(\begin{smallmatrix}
        x_1 & \theta_1 \\
        \theta_2 & x_2
    \end{smallmatrix}\right)
$ with multiplication
\begin{align} \label{eq:multipltable}
    \begin{pmatrix}
        x_1 & \theta_1 \\
        \theta_2 & x_2
    \end{pmatrix}
    \cdot
    \begin{pmatrix}
        y_1 & \xi_1 \\
        \xi_2 & y_2
    \end{pmatrix}
    =
    \begin{pmatrix}
        x_1 y_1+\theta_1\xi_2 & x_1\xi_1+ \theta_1 y_2 \\
        \theta_2 y_1 + x_2\xi_2 & x_2 y_2 + \theta_2\xi_1
    \end{pmatrix}
\end{align}
The corresponding reduced group is $\red{G} = \big(\R \setminus
\set{0}\big)^2$. A basis of left invariant vector fields $\gl(1|1)$
is easily recognized to be
\begin{align*}
    X_1 &= x_1 \frac{\partial}{\partial x_1} + \theta_2\frac{\partial}{\partial \theta_2} &
    X_2 &= x_2 \frac{\partial}{\partial x_2} + \theta_1\frac{\partial}{\partial \theta_1} \\
    T_1 &= x_1 \frac{\partial}{\partial \theta_1}-\theta_2\frac{\partial}{\partial x_2} &
    T_2 &= x_2 \frac{\partial}{\partial\theta_2}-\theta_ 1\frac{\partial}{\partial x_1}
\end{align*}
with commutation relations (for all $i,j = 1,2$)
\begin{align*}
    [X_i, X_j] &= 0 &
    [T_i, T_i] &= 0 \\
    [X_i, T_j] &= (-1)^{i+j} T_j &
    [T_1, T_2] &= -X_1-X_2
\end{align*}
$
    h = \left(\begin{smallmatrix}
        y_1 & 0 \\
        0 & y_2
    \end{smallmatrix}\right) \in \red{G}
$ acts through the adjoint representation on $\gl(1|1)_1$ as
follows:
\begin{align*}
    h.T_1 &= y_1 \, T_1 \, y_2^{-1} &
    h.T_2 &= y_2 \, T_2 \, y_1^{-1}
\end{align*}

Using the theory developed in the previous section, we now want to
reconstruct the multiplication map of $G$ in terms of the
corresponding SHCP. Introduce the linear operators
\begin{align*}
    \phi_i \colon \Lambda\big(\gl(1|1)_1\big) &\to \cinfty(\red{G}) \\
    1 &\mapsto y_i \\
    T_1, T_2, T_1 \wedge T_2 &\mapsto 0
\end{align*}
and
\begin{align*}
    \Phi_i \colon \Lambda\big(\gl(1|1)_1\big) &\to \cinfty(\red{G}) \\
    T_i &\mapsto 1 \\
    1,T_{j\neq i}, T_1 \wedge T_2 &\mapsto 0
\end{align*}
These are going to be our coordinates on $\HOM
\big(\Lambda\big(\gl(1|1)\big),\cinfty(\red{G})\big)$. These maps
extend in a natural way to  $\Uk(\gk_0)$-linear maps from
$\Uk(\gk_0)\otimes \Lambda(\gk_1)$ to $\cinfty(\red{G})$, which we
will denote by the same letter. We denote by $\widehat{\phi}$
(resp.\ $\widehat{\Phi}$) the composition $\phi \circ
\widehat{\gamma}^{-1}$ (resp.\ $\Phi \circ \widehat{\gamma}^{-1}$).

We want to calculate the pullbacks
\begin{align}
    \label{eq:exdef1}
    \begin{split}
        \big(\mu^\ast(\phi_i)\big)(X,Y)(g,h)
        &\coloneqq \widehat{\phi}_i \big( h^{-1}.{\gamma}(X) {\gamma}(Y) \big) (gh) \\
        &= \phi_i \Big( \widehat{\gamma}^{-1} \big( h^{-1}.{\gamma}(X) {\gamma}(Y) \big) \Big)(gh)
    \end{split} \\
    \label{eq:exdef2}
    \begin{split}
        \big(\mu^\ast(\Phi_i)\big)(X,Y)(g,h)
        &\coloneqq \widehat{\Phi}_i \big( h^{-1}.{\gamma}(X) {\gamma}(Y) \big) (gh) \\
        &= \Phi_i \Big( \widehat{\gamma}^{-1} \big( h^{-1}.{\gamma}(X) {\gamma}(Y) \big) \Big) (gh)
    \end{split}
\end{align}
In order to perform the computations we first need to compute the
elements $\widehat{\gamma}^{-1} \big( h^{-1}.{\gamma}(X) {\gamma}(Y)
\big)$. Next table collects them.
\medskip
\begin{center}
\begin{footnotesize}
\begin{tabular}{|c|c|c|c|c|}
    \hline
    \backslashbox{X}{Y} & $\boldsymbol{1}$ & $\boldsymbol{T_1}$ & $\boldsymbol{T_2}$ & $\boldsymbol{T_1 \wedge T_2}$ \\
    \hline
    $\boldsymbol{1}$ & $1$& $T_1$ & $T_2$ & $T_1 \wedge T_2$ \\
    \hline
    $\boldsymbol{T_1}$ & $ y_1^{-1}y_2 T_1$ & $0$ & $\begin{gathered} y_2^{-1}y_1\big(T_1\wedge T_2 \\ -\tfrac{1}{2}(X_1+X_2)\big)\end{gathered}$ & $ \frac{y_1^{-1}y_2}{2}(X_1 + X_2)T_1$ \\
    \hline
    $\boldsymbol{T_2}$ & $y_2^{-1}y_1 T_2 $ & $\begin{gathered}y_2^{-1}y_1\big(-T_1\wedge T_2 \\ -\tfrac{1}{2}(X_1+X_2)\big)\end{gathered}$ & $0$ & $ - \frac{y_2^{-1}y_1} {2}(X_1 + X_2) T_2$ \\
    \hline
    $\boldsymbol{T_1 \wedge T_2} $ & $T_1\wedge T_2$ &$ -\frac{1}{2}(X_1+X_2)T_1$ & $\frac{1}{2}(X_1+X_2)T_2$ &$ \frac{1}{4}(X_1+X_2)^2$ \\
    \hline
\end{tabular}
\end{footnotesize}
\end{center}
\medskip
From this and using definitions \eqref{eq:exdef1} and
\eqref{eq:exdef2}, we can calculate easily the various pullbacks.
Let us do it in detail in the case of $\phi_1$. In such a case the
pullback table of $\big(\mu^\ast(\phi_1)\big)(X,Y)
\big((x_1,x_2),(y_1,y_2)\big)$ is
\medskip
\begin{center}
\begin{tabular}{|c|c|c|c|c|}
    \hline
    \backslashbox{X}{Y} & $\boldsymbol{1}$ & $\boldsymbol{T_1}$ & $\boldsymbol{T_2}$ & $\boldsymbol{T_1 \wedge T_2}$ \\
    \hline
    $\boldsymbol{1}$ & $x_1 y_1$ & $0$ & $0$ & $0$ \\
    \hline
    $\boldsymbol{T_1}$ & $0$ & $0$ & $-\frac{1}{2}x_1 y_1$ & $0$ \\
    \hline
    $\boldsymbol{T_2}$ & $0$ & $-\frac{1}{2} y_2^{-1}x_1 y_1^2$ & $0$ & $0$ \\
    \hline
    $\boldsymbol{T_1 \wedge T_2} $ & $0$ &$ 0$ & $0$ &$ \frac{1}{4}x_1 y_1$ \\
    \hline
\end{tabular}
\end{center}
\medskip
The link with the form of the multiplication morphism as given in
eq.\ \eqref{eq:multipltable} is established by the isomorphism
\begin{align*}
    x_1 &= \phi_1\left(1 + \frac{\Phi_1\Phi_2}{2}\right) \\
    x_2 &= \phi_2\left(1 - \frac{\Phi_1\Phi_2}{2}\right) \\
    \theta_i &= \phi_i \Phi_i
\end{align*}
\end{example}

\section{$G$-supermanifolds} \label{sec:G-superman}

Let $M$ be a supermanifold and let $G$ denote a SLG with
multiplication, inverse and unit $\mu$, $i$ and $e$ respectively.

\begin{definition}
A morphism of supermanifolds
\[
    a \colon G\times M \to M
\]
is called an \emph{action} of $G$ on $M$ if it satisfies
\begin{subequations}
\begin{gather}
    \label{eq:action_mult}
    a \circ ( \mu \times \id_M ) = a \circ ( \id_G \times  a ) \\
    a \circ \langle \hat{e} , \id_M \rangle = \id_M
\end{gather}
\end{subequations}
(see footnote \ref{footnote:notations} for the notations).
\end{definition}

If an action $a$ of $G$ on $M$ is given, then we say that $G$ acts
on $M$, or that $M$ is a \emph{$G$-supermanifold}.

Using the functor of points language, an action of the SLG
$G{(\blank)}$ on the the supermanifold $M(\blank)$ is a natural
transformation
\[
    a(\blank) \colon G(\blank) \times M(\blank) \to M(\blank)
\]
such that, for each $S\in \smflds$, $a(S)\colon G(S)\times M(S)\to
M(S)$ is an action of  the set theoretical group $G(S)$ on the set
$M(S)$.

If $p \in \red{M}$ and $g \in \red{G}$,  we define for future use
the maps
\begin{subequations}
\begin{align}
    \label{eq:a_p}
    a_p \colon G &\to M &
    a_p &\coloneqq a \circ \langle \id_G , \hat{p} \rangle \\
    a^g \colon M &\to M &
    a^g &\coloneqq a \circ \langle \hat{g} , \id_M \rangle
\end{align}
\end{subequations}
that, in the functor of points notation, become
\begin{align*}
    &\begin{aligned}
        a_p(S) \colon G(S) &\to M(S) \\
        g &\mapsto g.\hat{p}
    \end{aligned} &
    &\begin{aligned}
        a^g(S) \colon M(S) &\to M(S) \\
        m &\mapsto \hat{g}.m
    \end{aligned}
\end{align*}
They obey the following relations
\begin{itemize}
    \item $a^g \circ a^{g^{-1}} = \id_M$ for all $g\in \red{G}$
    \item $a^g \circ a_p = a_p \circ \ell_g$ for all $g\in \red{G}$ and $p\in \red{M}$
\end{itemize}

As in the classical case the above relations play an important role
in proving that $a_p$ is a constant rank mapping (next lemma). In
the super context, this is a more delicate result than its classical
counterpart, since the concept of constant rank itself  is more
subtle (see \cite{Leites}). We briefly recall it. If
$
    M =
    \left(
    \begin{smallmatrix}
        A & B \\
        C & D
    \end{smallmatrix}
    \right)
$ is an even $p|q \times m|n$ matrix with entries in the sections
$\sheaf(U)$ over a superdomain $U$, then we say that $M$ has
constant rank $r|s$ if there exist $G_1 \in
\Gl_{p|q}\big(\sheaf(U)\big)$ and $G_2 \in
\Gl_{m|n}\big(\sheaf(U)\big)$ such that $G_1 M G_2$ has the form $
    \left(
    \begin{smallmatrix}
        A' & 0 \\
        0 & D'
    \end{smallmatrix}
    \right)
$ with $
    A' =
    \left(
    \begin{smallmatrix}
        \id_r & 0 \\
        0 & 0
    \end{smallmatrix}
    \right)
$ and $
    D' =
    \left(
    \begin{smallmatrix}
        \id_s & 0 \\
        0 & 0
    \end{smallmatrix}
    \right)
$. Finally, if $\psi \colon M \to N$ is a morphism between
supermanifolds, we say that $\psi$ has constant rank $r|s$ at $m \in
\red{M}$, if there exists a coordinate neighborhood of $m$ such that
the super Jacobian matrix $J_\psi$ has rank $r|s$. We can now prove
the following fundamental lemma.

\begin{lemma} \label{lemma:constant_rank}
$a_p$ has constant rank.
\end{lemma}

\begin{proof}
Let $\gk$ be the super Lie algebra of $G$ and let $J_{a_p}$ be the
Jacobian matrix of $a_p$. Since
\[
    \red{J_{a_p}}(g) = {(da_p)}_g = {(da^g)}_p {(da_p)}_e {(d\ell_{g^{-1}})}_g
\]
and $a^g$ and $\ell_{g^{-1}}$ are isomorphisms, $\red{J_{a_p}}(g)$
has rank $\dim \gk - \dim \ker {(da_p)}_e$ for each $g \in \red{G}$.
Moreover, recalling that, if $X \in \gk$, $D^L_X = (\id \otimes X)
\mu^*$ and using eq.\ \eqref{eq:action_mult} we have that, for each
$X \in \ker{(da_p)}_e$,
\begin{align*}
    D^L_X a_p^*
    &= (\id \otimes X) \mu^* (\id \otimes \ev_p) a^* \\
    &= (\id \otimes X \otimes \ev_p) (\id \otimes \mu^*) a^* \\
    &= \big( \id \otimes {(da_p)}_e(X) \big) a^* = 0
\end{align*}

If $\set{x_i,\theta_j}$ and $\set{y_k,\xi_l}$ are coordinates
 in a neighbourhood $U$ of $e$, and in a neighbourhood $V
\supseteq \red{a_p}(U)$ of $p$ respectively, then
\[
    J_{a_p} =
    \begin{pmatrix}
        \frac{\partial a_p^*(y_k)}{\partial x_i} & - \frac{\partial a_p^*(y_k)}{\partial \theta_j} \\
        \frac{\partial a_p^*(\xi_l)}{\partial x_i} & \frac{\partial a_p^*(\xi_l)}{\partial \theta_j}
    \end{pmatrix}
\]
Let $m|n = \dim \ker {(da_p)}_e$ and let $\set{X_u}$ and $\set{T_v}$
be bases of $\gk_0$ and $\gk_1$ such that $X_u,T_v \in \ker
{(da_p)}_e$ for $u \leq m$ and $v \leq n$. If
\begin{align*}
    D^L_{X_u} &= \sum_i a_{u,i} \frac{\partial}{\partial x_i} + \sum_j b_{u,j} \frac{\partial}{\partial \theta_j} &
    D^L_{T_v} &= \sum_i c_{v,i} \frac{\partial}{\partial x_i} + \sum_j d_{v,j} \frac{\partial}{\partial \theta_j}
\end{align*}
(with $a_{u,i},d_{v,j} \in \sheaf_G(U)_0$ and $b_{u,j},c_{v,i} \in
\sheaf_G(U)_1$) and $A = \left( \begin{smallmatrix} a_{u,i} &
-c_{v,i} \\ b_{u,j} & d_{v,j} \end{smallmatrix} \right)$, then the
matrix
\[
    J_{a_p} A =
    \begin{pmatrix}
        D^L_{X_u} a_p^*(y_k) & - D^L_{T_v} a_p^*(y_k) \\
        D^L_{X_u} a_p^*(\xi_l) & D^L_{T_v} a_p^*(\xi_l)
    \end{pmatrix}
\]
has $m+n$ zero columns. Since $\set{X_u,T_v}$ is a  linearly
independent set of vectors, $A$ is invertible and so, for
\cite[lemma~2.3.8]{Leites} $J_{a_p}$ has constant rank in $U$ and,
by translation, in all $\red{G}$.
\end{proof}

Since the category of SLG is equivalent to the category of SHCP, one
could ask whether there is an equivalent notion of action of a SHCP
on a supermanifold. The answer is affirmative and it is given in the
next proposition (see also \cite{DM}).

\begin{proposition} \label{prop:action_SLG_SHCP}
Suppose $G$ acts on a supermanifold $M$, then there are
\begin{enumerate}
    \item an action
    \begin{equation} \label{eq:interesting}
        \ua \colon \red{G} \times M \to M
    \end{equation}
    $\ua \coloneqq a \circ (j_{\red{G}\to G} \times
    \id_M)$ of the reduced Lie group $\red{G}$ on the supermanifold
    $M$;
    \item a representation
        \begin{equation} \label{eq:infinitesimalaction}
        \begin{aligned}
            \rho_a \colon \gk &\to \VEC(M)\op \\
            X &\mapsto \left( X \otimes \id_{\sheaf(M)} \right) a^\ast
        \end{aligned}
    \end{equation}
    of the super Lie algebra $\gk$ of $G$ on the opposite of the Lie
    algebra of vector fields over $M$.
\end{enumerate}
The above two maps satisfy the following compatibility relations
\begin{subequations} \label{eqs:compatibilityforactions}
\begin{align}
    \restr{\rho_a}{\gk_0}(X) &= \left( X \otimes \id_{\sheaf(M)} \right) \ua^\ast &
    &\forall X \in \gk_0 \\
    \rho_a(g.Y) &= \big(\ua^{g^{-1}}\big)^\ast \rho_a(Y) {(\ua^g)}^\ast &
    &\forall g \in \red{G}, \, Y \in \gk
\end{align}
\end{subequations}
Conversely, let $(\red{G},\gk)$ be the SHCP associated with $G$ and
let maps $\ua$ and $\rho$ like in points 1 and 2 above satisfying
conditions \eqref{eqs:compatibilityforactions} be given. There is a
unique action $a_\rho \colon G\times M \to M$ of the SLG $G$ on $M$
whose reduced and infinitesimal actions are the given ones. It is
given by
\begin{equation} \label{eq:reconstructingactions}
    \begin{aligned}
        a_\rho^* \colon \sheaf(M) &\to \HOM_{\Uk(\gk_0)} \big( \Uk(\gk), \cinfty(\red{G}) \hotimes \sheaf(M) \big) \\
        f &\mapsto \Big[ X \mapsto (-1)^{\p{X}} \big( \id_{\cinfty(\red{G})} \otimes \rho(X) \big) \ua^\ast(f) \Big]
    \end{aligned}
\end{equation}
\end{proposition}

In analogy with the classical case, one can use super Frobenius
theorem to reconstruct a local action from an infinitesimal action
\eqref{eq:infinitesimalaction}. Nevertheless it is particularly
interesting that the assignment of \eqref{eq:interesting} allows to
avoid the use of super Frobenius theorem and makes possible an
explicit reconstruction of the global action. The form of the
reconstruction formula given by eq.\
\eqref{eq:reconstructingactions} can be easily obtained as follows.

Let $a$ be an  action of $G$ on $M$ and let $(\ua,\rho_a)$ be as in
prop.\ \ref{prop:action_SLG_SHCP}. If $f \in \sheaf(M)$, then
\[
    a^\ast(f) \in \HOM_{\Uk(\gk_0)} \big( \Uk(\gk), \cinfty(\red{G}) \big) \hotimes \sheaf(M)
    \isom \HOM_{\Uk(\gk_0)} \big( \Uk(\gk), \cinfty(\red{G}) \hotimes \sheaf(M) \big)
\]
hence, using eq.\ \eqref{eq:action_mult} and the fact that $\rho_a$
is an antihomomorphism, for all $X \in \Uk(\gk)$
\begin{align*}
    a^\ast(f)(X)
    &= (-1)^{\p{X}} \big[ (D^L_X \otimes \id) a^\ast(\phi) \big](1) \\
    &= (-1)^{\p{X}} \big( \id \otimes \rho_a(X) \big) \big( a^\ast(f)(1) \big) \\
    &= (-1)^{\p{X}} \big( \id \otimes \rho_a(X) \big) \ua^\ast(f)
\end{align*}
This also proves the uniqueness statement of the theorem and it
suggests how to prove the existence of the action $a_\rho$.

\begin{proof}[Proof of prop.\ \ref{prop:action_SLG_SHCP}]
Let us check that $a_\rho^*(f)$ is $\Uk(\gk_0)$-linear. For all $X
\in \Uk(\gk)$ and $Z \in \gk_0$ we have
\begin{align*}
    a_\rho^*(f)(ZX)
    &= (-1)^{\p{X}} \big( \id \otimes \rho(ZX) \big) \ua^*(f) \\
    &= (-1)^{\p{X}} \big( \id \otimes \rho(X) \big) (\id \otimes Z_e \otimes \id) (\id \otimes \ua^*) \ua^*(f) \\
    &= (-1)^{\p{X}} \big( \id \otimes \rho(X) \big) (\id \otimes Z_e \otimes \id) (\red{\mu}^* \otimes \id) \ua^*(f) \\
    &= \big( \red{D^L_Z} \otimes \id \big) \big[ a_\rho^*(f)(X) \big]
\end{align*}
We now check that $a_\rho^*$ is a superalgebra morphism.
\begin{align*}
    \big[ a_\rho^*(f_1) \cdot a_\rho^*(f_2) \big](X)
    &= m_{\cinfty(\red{G}) \hotimes \sheaf(M)} \big[ a^*(f_1) \otimes a^*(f_2) \big] \Delta(X) \\
    &= (-1)^{\p{X}} m \Big[ \big(\id \otimes \rho(X_{(1)})\big) \ua^*(f_1) \otimes \big(\id \otimes \rho(X_{(2)})\big) \ua^*(f_2) \Big] \\
    &= (-1)^{\p{X}} \big( \id \otimes \rho(X) \big) \big( \ua^*(f_1) \cdot \ua^*(f_2) \big) \\
    &= a_\rho^*(f_1 \cdot f_2)(X)
\end{align*}
where $f_i \in \sheaf(M)$ and $X_{(1)} \otimes X_{(2)}$ denotes
$\Delta(X)$. Concerning the ``associative'' property, we have that,
for $X,Y \in \Uk(\gk)$ and $g,h \in \red{G}$,
\begin{align*}
    \big[ (\mu^* \otimes \id) a_\rho^*(f) \big](X,Y)(g,h)
    &= \big[ a_\rho^*(f) \big] (h^{-1}.X Y)(gh) \\
    &= (-1)^{\p{X} + \p{Y} + \p{X}\p{Y}} \rho(Y) \rho(h^{-1}.X) {(\ua^{gh})}^*(f) \\
    &= (-1)^{\p{X} + \p{Y} + \p{X}\p{Y}} \rho(Y) {(\ua^h)}^* \rho(X) {(\ua^g)}^*(f) \\
    &= \big[ (\id \otimes a_\rho^*) a_\rho^*(f) \big](X,Y)(g,h)
\end{align*}
and, finally, $(\ev_e \otimes \id)a_\rho^*(f) = \rho(1)
{(\ua^e)}^*(f) = f$.
\end{proof}

We end this section resuming  example \ref{example:GL(1|1)}.

\begin{example}
Consider again  the SLG $G = \Gl(1|1)$ introduced in example
\ref{example:GL(1|1)}. $G$ acts on itself by left multiplication,
and, using the same notations as in the previous example,  we have
\begin{enumerate}
    \item left action of $\red{G}$ on $G$:
    \[
        \begin{pmatrix}
            x_1 & 0\\
            0 & x_2
        \end{pmatrix}
        \cdot
        \begin{pmatrix}
            y_1 & \xi_1\\
            \xi_2 & y_2
        \end{pmatrix}
        =
        \begin{pmatrix}
            x_1y_1 & x_1\xi_1\\
            x_2\xi_2 & x_2 y_2
        \end{pmatrix}
    \]
    \item representation of $\gl(1|1)$ on the super Lie algebra
    $\VEC(G)\op$:
    \begin{align*}
        X_1 &\mapsto y_1\frac{\partial}{\partial y_1}+\xi_1 \frac{\partial}{\partial \xi_1} &
        X_2 &\mapsto y_2\frac{\partial}{\partial y_2}+\xi_2 \frac{\partial}{\partial \xi_2} \\
        T_1 &\mapsto y_2\frac{\partial}{\partial \xi_1}+\xi_2 \frac{\partial}{\partial y_1} &
        T_2 &\mapsto y_1\frac{\partial}{\partial \xi_2}+\xi_1 \frac{\partial}{\partial y_2}
    \end{align*}
    In this case the representation sends each element of $\gl(1|1)$
    into the corresponding  right invariant vector field.
\end{enumerate}
The action $\mu$ can be reconstructed using eq.\
\eqref{eq:reconstructingactions}; a simple calculation shows that
\begin{align*}
    \mu^\ast(x_1) &=  x_1 y_1(1 +\theta_1 \theta_2) + x_1 \xi_2\theta_1 &
    \mu^\ast(x_2) &= x_2 y_2 (1 +\theta_1 \theta_2) + x_2 \xi_1\theta_2 \\
    \mu^\ast(\theta_1) &=  x_1\xi_1 (1 +\theta_1\theta_2) + x_1 y_2 \theta_1 &
    \mu^\ast(\theta_2) &= x_2\xi_2 (1 -\theta_1\theta_2) + x_2 y_1 \theta_2
\end{align*}
The usual form of the multiplication map (as given in example
\ref{example:GL(1|1)}) is obtained using the isomorphism
\begin{align*}
    x_1 &\mapsto x_1 (1+\theta_1\theta_2) &
    x_2 &\mapsto x_2 (1+\theta_1\theta_2) \\
    \theta_1 &\mapsto \frac{\theta_1}{x_1} &
    \theta_2 &\mapsto \frac{\theta_2}{x_2}
\end{align*}
\end{example}

\section{Transitive actions} \label{sec:trans_action}

Let $M$ be a $G$-supermanifold with respect to an action $a \colon G
\times M \to M$. Next definition is the natural generalization of
the classical one.

\begin{definition}
Suppose $G$ acts on $M$ through $a \colon G\times M \to M$. We say
that $a$ is \emph{transitive} if there exists $p \in \red{M}$ such
that $a_p$ (see eq.\ \eqref{eq:a_p}) is a surjective submersion.
\end{definition}

\begin{remark}
Since $a_{g.p}= a_p \circ r_g$, if $a_p$ is submersive for one $p
\in \red{M}$ then it is submersive for all $p \in \red{M}$.
\end{remark}

Next proposition characterizes transitive actions.

\begin{proposition}
Suppose $M$ is a  $G$-superspace, then the following facts are
equivalent:
\begin{enumerate}
    \item $M$ is transitive;
    \item
    \begin{itemize}
        \item $\red{a} \colon \red{G} \times \red{M} \to \red{M}$ is
        transitive;
        \item ${(da_p)}_e \colon \gk \to T_p(M)$ is
        surjective;
    \end{itemize}
    \item if $q$ denotes the odd dimension of $G$, then
    \[
        a_p(\R^{0|q}) \colon G(\R^{0|q}) \to M(\R^{0|q})
    \]
    is surjective;
    \item\label{it:sheafific} the sheafification of the functor
    \[
        \begin{aligned}
            \smflds\op &\to \sets \\
            S &\mapsto (\im a_p)(S) \coloneqq \Set{a_p \circ \phi | \phi \in G(S)}
        \end{aligned}
    \]
    is the functor of points of $M$.
\end{enumerate}
\end{proposition}

\begin{proof}
The second statement is an immediate consequence of lemma
\ref{lemma:constant_rank} and previous remark.

 Let us hence check
the equivalence of the third with the first. If $\phi \in
M(\R^{0|q}) = \Hom(\R^{0|q},M)$, let $\red{\phi} \in \red{M}$ be the
image of the reduced map associated with $\phi$. It is clear that
the pullback $\phi^*$ depends only on the restriction of the
sections of $\sheaf_M$ to an arbitrary neighbourhood of
$\red{\phi}$. If $a_p$ is a surjective submersion, there exists a
local right inverse $s$ of $a_p$ defined in a neighbourhood of
$\red{\phi}$. By the locality of $\phi$, $s \circ \phi$ is a well
defined element of $G(\R^{0|q})$ and moreover
\[
    \big[ a_p(\R^{0|q}) \big](s \circ \phi) = a_p \circ s \circ \phi = \phi
\]
so that $a_p(\R^{0|q})$ is surjective.

Suppose, conversely, $a_p(\R^{0|q})$ surjective. Looking at the
reduced part of each morphism in $a_p(\R^{0|q})\big(
G(\R^{0|q})\big)$, we have that $a_p(\R^{0|0}) = \red{a_p} \colon
\red{G} \to \red{M}$ is surjective. As a consequence (see
\cite[th.~5.14]{Michor}), $\red{a}$ is a classical transitive action
and $\red{a}_p$ is a submersion. Let now $m \in \red{M}$ and
$\set{x_i,\theta_j}$ be coordinates in a neighbourhood $U$ of it.
Consider the following element of $M(\R^{0|q})$ defined by the
pullback
\[
    \begin{aligned}
        \phi^* \colon \sheaf_M(U) &\to \sheaf(\R^{0|q}) = \Lambda(\eta_1,\ldots,\eta_q) \\
        x_i &\mapsto \red{x_i}(m) \\
        \theta_j &\mapsto \eta_j
    \end{aligned}
\]
By surjectivity of $a_p(\R^{0|q})$, there exists $\psi \in
G(\R^{0|q})$ such that
\begin{align*}
    \psi^\ast \circ a^\ast_p (x_i) &= \red{x_i}(m) \\
    \psi^\ast \circ a^\ast_p (\theta_j) &= \eta_j
\end{align*}
and this implies that $T_m(M)_1$ is in the image of
${(da_p)}_{\red{\psi}}$. Since, by previous considerations,
$\red{a_p}$ is a submersion, also $T_m(M)_0$ is in the image. Hence,
due to lemma \ref{lemma:constant_rank}, we are done. For the last
point see \cite{BCF}.
\end{proof}

\begin{remark}
In the third point of the above proposition, it is not possible to
require the transitivity of $a(S)$ for each $S$. Indeed, in such a
case, each map $S \to M$ can be lifted to a map $S \to G$. This in
particular implies  the existence of a global section of the
fibration $G \to M$ (take $S=M$ and the identity map). This problem
is solved exactly by taking the sheafification of the functor as
indicated in in point \ref{it:sheafific}.
\end{remark}

\section{Stabilizer} \label{sec:stabilizer}

Let $G$ be a SLG, and suppose $M$ is a $G$-supermanifold; the aim of
this section is to define the notion of stability subgroup and to
characterize it from different perspectives.

We start recalling the definition of an equalizer. Given two objects
($X$ and $Y$) and two arrows ($\alpha$ and $\beta$) between them, an
equalizer is a universal pair $(E, \epsilon)$ that makes
\[
    \xymatrix@1{E \ar[r]^\epsilon & X \ar@<.7ex>[r]^\alpha \ar@<-.7ex>[r]_\beta & Y}
\]
commuting. This means that  if $\tau \colon T \to X$ is such that
$\alpha \circ \tau = \beta \circ \tau$, then there exists a unique
$\sigma \colon T \to E$ such that $\epsilon \circ \sigma = \tau$. If
an equalizer exists, it is unique up to isomorphism.

One can easily convince himself that the next definition mimic the
classical one.

\begin{definition}
Suppose $G$ is a SLG and $a\colon G\times M \to M$ is an action of
$G$ on the supermanifold $M$. We call \emph{stabilizer} of  $p\in
\red{M}$ the supermanifold $G_p$ equalizing the diagram (where
$\hat{p} \colon G \to M$ is as in footnote \ref{footnote:notations})
 \[
        \xymatrix@1{ G \ar@<.7ex>[r]^{a_p} \ar@<-.7ex>[r]_{\hat{p}} & M}
    \]
\end{definition}

It is not a priori obvious that such an equalizer exists. Next
proposition shows that the definition is meaningful and
characterizes the notion  of stabilizer both from the point of view
of the functor of points and in terms of the corresponding SHCP.

\begin{proposition}
\label{prop:SHCP_stab}
\begin{enumerate}
    \item The diagram
    \[
        \xymatrix@1{ G \ar@<.7ex>[r]^{a_p} \ar@<-.7ex>[r]_{\hat{p}} & M}
    \]
    admits an equalizer $G_p$;
    \item $G_p$ is a sub-SLG of $G$;
    \item the functor $S \mapsto G(S)_{\hat{p}}$ assigning  to each supermanifold
    $S$ the stabilizer of $\hat{p}$ of the action of $G(S)$ on $M(S)$ is
    represented by $G_p$;
    \item let $(\red{G}_p, \gk_p)$ be the SHCP associated with the stabilizer
    $G_p$. Then $\red{G}_p \subseteq \red{G}$ is the classical stabilizer
    of $p$ with respect to the reduced action and $\gk_p = \ker da_p$.
\end{enumerate}
\end{proposition}

\begin{proof}
Let us put ourself in a general context and let us suppose that
$\psi \colon M \to N$ is a morphism of constant rank between two
supermanifold. If $p \in \red{N}$, define $\sheaf[I]_p = \set{f \in
\sheaf(N) | \red{f}(p) = 0}$ and $\sheaf[J]_p^\psi$ as the ideal in
$\sheaf(M)$ generated by $\psi^*(\sheaf[I]_p)$. In this case there
is a unique closed subsupermanifold $S$ of $M$ such that
$\sheaf[J]_p^\psi = \ker j_S^*$, where $j_S^*$ is the pullback of
the embedding $j_S \colon S \to M$. This submanifold is denoted with
$\psi^{-1}(p)$ (see \cite[\S~3.2.9]{Leites}). Let $G_p$ be
$a_p^{-1}(p)$. We are going to see that, as in classical context,
$G_p$ is the stabilizer of $p$.

First of all we recall that, if $A$ and $B$ are two generic algebras
and $\alpha$ and $\beta$ are morphisms between them, as it is easy
to check, their coequalizer --- the equalizer in the opposite
category --- is the algebra $C = B/J$, where $J = \langle\,
\alpha(a) - \beta(a) \,|\, a \in A \,\rangle$ is the ideal generated
by $\set{\alpha(a) - \beta(a) | a \in A}$.

Since the embedding $j_{G_p} \colon G_p \to G$ is regular and
closed, $j_{G_p}^*$ is surjective (see \cite[\S~3.2.5]{Leites}).
Hence $\sheaf(G_p) \isom \sheaf(G)/\ker j_{G_p}^*$, and moreover
\begin{multline*}
    \ker j_{G_p}^*
    = \left\langle\, a_p^*(f) \,\middle|\, f \in \sheaf[I]_p \,\right\rangle = \\
    = \left\langle\, a_p^*\big( f - \red{f}(p) \big) \,\middle|\, f \in \sheaf(M) \,\right\rangle
    = \left\langle\, a_p^*(f) - \hat{p}^*(f) \,\middle|\, f \in \sheaf(M) \,\right\rangle
\end{multline*}
Therefore $\sheaf(G_p)$ is the coequalizer of
\[
    \xymatrix@1{\sheaf(M) \ar@<.7ex>[r]^{a_p^*} \ar@<-.7ex>[r]_{\hat{p}^*} & \sheaf(G) \ar[r]^{j_{G_p}^*} & \sheaf(G_p)}
\]
and hence $G_p$ is the equalizer of
\[
    \xymatrix@1{G_p \ar[r]^{j_{G_p}} & G \ar@<.7ex>[r]^{a_p} \ar@<-.7ex>[r]_{\hat{p}} & M}
\]

This concludes item 1. In order to prove point 2 we have to show
that $G_p$ is a sub-SLG of $G$. Due to Yoneda lemma, if we prove
item 3 also 2 is done. On the other hand item 3 can be proved easily
noticing that the functor $G(\blank)_{\hat{p}}$ equalizes the
natural transformations
\[
    \xymatrix@1@C=8ex{G(\blank) \ar@<.7ex>[r]^{a_p(\blank)} \ar@<-.7ex>[r]_{\hat{p}(\blank)} & M(\blank)}
\]
and, since the Yoneda embedding preserves equalizers and due to
their uniqueness, $G_p(\blank) \isom G(\blank)_{\hat{p}}$.

Let us finally consider item 4. The first statement is clear since
$\red{G} \isom G(\R^{0|0})$ as set theoretical groups. Moreover,
since, for all $f \in \sheaf(M)$, $j_{G_p}^* \circ a_p^*(f)$ is a
constant, $\gk_p \subseteq \ker da_p$ and they are equal for
dimension considerations.
\end{proof}

\section{Homogeneous supermanifolds}

In this section we give a detailed account of homogeneous
supermanifolds. Essentially all the results presented in this
section are known (see \cite{Kostant}, and \cite{FLV}). We
nevertheless spend some time in proving them since we adopt a
slightly different approach through Koszul's realization of the
sheaf. This allows us to give a very explicit description of the
structure sheaf of the homogeneous supermanifold  (lemma
\ref{lemma:OGH_OHG} below) and to prove its local triviality without
using super Frobenius theorem (proposition \ref{prop:localtriv}
below). On the other hand  proposition \ref{prop:uniqueness}
completely relies on \cite{FLV}.

Let $G$ be a SLG and let $H$ be a closed sub-SLG. Let $\gk$ and
$\hk$ be the respective super Lie algebras. Let $U\subseteq
\red{G}/\red{H}$ and $V\subseteq \red{H}\backslash \red{G}$ be open
sets and define
\begin{align}
    \sheaf_{G/H}(U)
    &\coloneqq \Set{ \phi \in \sheaf_G \big( \pi^{-1}(U) \big) |
    \begin{array}{@{}cc@{}}
        r_h^\ast(\phi) = \phi & \forall h \in \red{H} \\
        D^L_X\phi=0 & \forall X\in \hk
    \end{array}
    } \\
    \label{eq:OHG}
    \sheaf_{H\backslash G}(V)
    &\coloneqq \Set{ \phi \in \sheaf_G \big( \pi^{-1}(V) \big) |
    \begin{array}{@{}cc@{}}
        \ell_h^\ast(\phi) = \phi & \forall h \in \red{H} \\
        D^R_X \phi =0 & \forall X \in \hk
    \end{array}
    }
\end{align}
where $\pi$ denotes, for simplicity, both the projections $\red{G}
\to \red{G}/\red{H}$ and $\red{G} \to \red{H}\backslash\red{G}$.
Define now the morphisms
\begin{align*}
    \mu_{G,H} &\colon G \times H \xrightarrow{\id_G \times j_{H \to G}} G \times G \stackrel{\mu}{\to} G \\
    \mu_{H,G} &\colon H \times G \xrightarrow{j_{H \to G} \times \id_G} G \times G \stackrel{\mu}{\to} G
\end{align*}
Next lemma shows that $\sheaf_{G/H}(U)$ (resp.\ $\sheaf_{H\backslash
G}(V)$) can be interpreted as the set of sections over $\pi^{-1}(U)$
(resp.\ $\pi^{-1}(V)$) that are right (resp.\ left) $H$-invariant.

\begin{lemma} \label{lemma:OGH_OHG}
$\sheaf_{G/H}$ and $\sheaf_{H\backslash G}$ define sheaves of
superalgebras over $\red{G}/\red{H}$ and $\red{H}\backslash\red{G}$
respectively. Moreover
\begin{align*}
    \sheaf_{G/H}(U) &= \Set{ \phi \in \sheaf_G \big( \pi^{-1}(U) \big) | \mu_{G,H}^\ast(\phi) = \pr_1^\ast(\phi) } \\
    \sheaf_{H\backslash G}(V) &= \Set{ \phi \in \sheaf_G \big( \pi^{-1}(V) \big) | \mu_{H,G}^\ast(\phi) = \pr_2^\ast(\phi) }
\end{align*}
where $\pr_i$ is the projection into the $i^\text{th}$ factor.
\end{lemma}

\begin{proof}
The first statement is easy to establish. We only show the first
equality, the proof of the second being equal.

Suppose $\phi$ belongs to the set in the r.~h.~s., then
\begin{gather*}
    r^\ast_h(\phi) = (\id \otimes \ev_h) \mu_{G,H}^\ast(\phi) = (\id \otimes \ev_h) \pr_1^\ast(\phi) = \phi \\
    D^L_X \phi = (\id \otimes X) \mu_{G,H}^\ast(\phi) = (\id \otimes X)(\phi \otimes 1) = 0
\end{gather*}
Conversely, suppose $\phi \in \sheaf_{G/H}(U)$, $X \in \Uk(\gk)$, $Y
\in \Uk(\hk)$, $g \in \red{G}$ and $h \in \red{H}$, then
\begin{align*}
    \big[ \big( \mu_{G,H}^\ast(\phi) \big)(X,Y) \big] (g,h)
    &= \big[ \phi(h^{-1}.X Y) \big](gh) \\
    &= (-1)^{\p{Y}} \big[ r_h^\ast(D^L_Y \phi)(X) \big](g) \\
    &= \begin{cases}
        0 & \text{if $Y \not\in \R$} \\
        \big[ \phi(X) \big](g) & \text{if $Y=1$}
    \end{cases} \\
    &= \big[ \big(\pr_1^\ast(\phi) \big)(X,Y) \big](g,h)
    \qedhere
\end{align*}
\end{proof}

\begin{proposition}
\label{prop:localtriv} $G/H \coloneqq
(\red{G}/\red{H},\sheaf_{G/H})$ and $H\backslash G \coloneqq
(\red{H}\backslash\red{G},\sheaf_{H\backslash G})$ are isomorphic
supermanifolds.
\end{proposition}

\begin{proof}
We start by showing that $G/H$ is a supermanifold. In view of lemma
\ref{lemma:OGH_OHG}, it only remains to prove the local triviality
of the sheaf.

Let $\hk$ be the super Lie algebra of $H$ and let $\gk = \hk \oplus
\pk$ be a homogeneous decomposition. Moreover let $s \colon U \to
\red{G}$, $U \subseteq \red{G}/\red{H}$, be a local section in a
neighbourhood of $\dot{e}$ (the equivalence class of $e$). Consider
the trivializing map
\begin{equation} \label{eq:trivhomog}
    \begin{aligned}
        \sheaf_{G/H}(U) &\to \HOM \big( \Lambda(\pk_1), \cinfty_{\red{G}/\red{H}}(U) \big) \\
        \phi &\mapsto \overline{\phi}
    \end{aligned}
\end{equation}
where, if $P \in \Lambda(\pk_1)$, $\dot{g} \in U$ and $\gamma$ is as
in lemma \ref{lemma:antisym},
\[
    \big[ \overline{\phi}(P) \big](\dot{g}) \coloneqq \big[\phi\big( \gamma(P) \big)\big]\big( s(\dot{g}) \big)
\]
The bijectivity of this map can be obtained easily from the
following remarks:
\begin{itemize}
    \item each $g \in \pi^{-1}(U)$ can be written uniquely as $g =
    s(\dot{g})h(g)$, with $h(g) \in \red{H}$;
    \item due to lemma \ref{lemma:antisym} each $X \in \Uk(\gk)$ can be
    written as $X = X_0 \gamma(X_1)$ with $X_0 \in
    \Uk(\gk_0)$ and $X_1 \in \Lambda(\gk_1)$;
    \item due to $\Uk(\gk_0)$-linearity and condition
    $r_h^* \phi = \phi$ for each $h \in \red{H}$, each
    $\phi \in \sheaf_{G/H}(U)$ satisfies
    \begin{align*}
        \big[ \phi(X) \big](g)
        &= \big[ \phi \big( X_0 \gamma(X_1) \big)\big] \big( s(\dot{g}) h(g) \big) \\
        &= \left[ \red{D^L_{h(g).X_0}} \phi \big( h(g).\gamma(X_1) \big)\right] \big( s(\dot{g}) \big)
    \end{align*}
    \item due to condition $D^L_H \phi = 0$ for each $H \in \hk_1$,
    $\phi$ is determined by its value on
    $\gamma\big(\Lambda(\pk_1)\big)$; indeed, if $H \in \hk_1$ and
    $X_i \in \gk_1$,
    \begin{align*}
        D^L_H \phi \big( \gamma(X_1 \wedge \dots \wedge X_n) \big)
        = - \phi \big( \gamma(X_1 \wedge \dots \wedge X_n \wedge H) \big) + \phi(Y) = 0
    \end{align*}
    with
    \[
        Y \in (\R \oplus \gk_0) \gamma \left( \bigoplus_{i=0}^{n-1} \Lambda^i(\gk_1) \right)
    \]
    (see \cite[lemma~2.3]{Koszul}) and than $\phi$ can be calculated by
    induction on $n$ once it is known
    on $\gamma\big(\Lambda(\pk_1)\big)$;
    \item the condition $D^L_H \phi = 0$ for each $H \in \hk_0$ does
    not give further restriction since $D^L_H = \frac{d}{dt} r_{\exp tH}^*
    \big|_{t=0}$.
\end{itemize}
Moreover, since $(\gamma \otimes \gamma) \circ
\Delta_{\Lambda(\pk_1)} = \Delta_{\Uk(\gk)} \circ
\restr{\gamma}{\Lambda(\pk_1)}$, the maps in eq.\
\eqref{eq:trivhomog} is easily seen to be a superalgebra morphism.
To end the first part of the proof notice that each $g\in \red{G}$
acts by left translation on the sheaf $\sheaf_G$ and that its action
preserves $\sheaf_{G/H}$. Hence $g$ acts as an algebra isomorphism
on $\sheaf_{G/H}$, so that local triviality is proved at all points
of $\red{G}/\red{H}$.

For $H \backslash G$ the proof is analogous. As in the classical
case, in order to show that $G/H$ and $H\backslash G$ are isomorphic
supermanifolds it is enough to consider the inverse morphism $i$
defined by \eqref{eq:inv}. Notice indeed that  $i^\ast$ sends
$\sheaf_{G/H}$ to $\sheaf_{H\backslash G}$ (and vice versa).
\end{proof}

\begin{definition}
We call $G/H$ (resp.\ $H\backslash G$) the \emph{homogeneous
supermanifold} of left (resp.\ right) invariant cosets.
\end{definition}

Next proposition establishes some properties of the manifold $G/H$.

\begin{proposition}
\label{prop:uniqueness} $G/H$ has the following properties
\begin{enumerate}
    \item $\pi \colon G \to G/H$ is a submersion whose reduced part
    $\red{\pi} \colon \red{G} \to \red{G}/\red{H}$ is the natural
    projection;
    \item there is a unique action $\beta$ of $G$ on $G/H$ such that the following diagram is commutative
    \[
        \xymatrix@C=8ex{
            G \times G \ar[r]^\mu \ar[d]_{\id_G \times \pi} & G\ar[d]^{\pi} \\
            G \times G/H \ar[r]^\beta & G/H
        }
    \]
\end{enumerate}
If a supermanifold $X$ exists satisfying the above properties then
$X \isom G/H$.
\end{proposition}

\begin{proof}
Essentially all assertions are consequences of the fact that
\[
    \begin{aligned}
        \pi^\ast \colon \sheaf_{G/H} &\to \sheaf_G \\
        f &\mapsto f
    \end{aligned}
\]
For all details, see \cite{FLV}.
\end{proof}

Next proposition proves that, exactly as in the classical case, each
transitive supermanifold is isomorphic to a homogeneous
supermanifold (see \cite{Kostant}).

\begin{proposition}
Let $M$ be a transitive $G$-supermanifold. If $p\in \red{M}$ and
$G_p$ is the stabilizer  of $p$, then there exists a $G$-equivariant
isomorphism
\[
    G/G_p \to M
\]
\end{proposition}

\begin{proof}
Fix a point $p\in \red{M}$ and consider the map $a_p\colon G \to M$.
If $U \subseteq \red{M}$, notice that $a^\ast_p\big(\sheaf_M(U)\big)
\subseteq \sheaf_{G/G_p}\big(\red{a_p^{-1}}(U)\big)$. Indeed, if $h
\in \red{G_p}$ and $X \in \gk_p = \Lie(G_p)$, for each $f \in
\sheaf_M(U)$
\[
    r_h^\ast \big(a_p^\ast(f)\big)
    = a^\ast_{h.p}(f) = a^\ast_p(f)
\]
and, due to prop.\ \ref{prop:SHCP_stab},
\begin{align*}
    D^L_X \big(a_p^\ast(f)\big)
    &= (\id \otimes X) \mu^\ast (\id \otimes \ev_p) a^\ast(f) \\
    &= (\id \otimes X) (\id \otimes a_p^\ast) a^\ast(f) \\
    &= \big( \id \otimes {(da_p)}_e(X) \big) a^\ast(f) = 0
\end{align*}
Hence we can define a map
\[
    \eta \colon G/G_p \to M
\]
through $\eta^\ast\coloneqq a^\ast_p$. It is easy to see that such a
map is $G$-equivariant:
\[
    a \circ (\id_G \times \eta) = \eta \circ \beta
\]
Finally, since $\red{\eta}$ is bijective and $d\eta$ is bijective at
each point ($da_p$ is surjective for transitivity hypothesis and
$\gk_p = \ker da_p$, so $d\eta$ is bijective at $\dot{e}$ and at
each point because of the equivariance), $\eta$ is an isomorphism
(see corollary to th.\ 2.16 in \cite{Kostant}).
\end{proof}

\paragraph{Acknowledgements.}
We deeply thanks Professor V.~S.~Varadarajan for having taught us so
much during the years, for the privilege of his collaboration and,
most of all, for his friendship. In particular our interest in super
Lie groups and their representations has its first origin in the
lectures he gave in Genoa, and our work on this subject has always
been done with his collaboration, his advice and his encouragement.

\end{document}